\documentclass[conference]{IEEEtran}
\IEEEoverridecommandlockouts
\usepackage{cite}
\usepackage{amsmath,amssymb,amsfonts}
\usepackage{algorithmic}
\usepackage{graphicx}
\usepackage{textcomp}
\usepackage{xcolor}
\usepackage{booktabs}   
\usepackage{tabularx}   
\usepackage{subcaption}
\usepackage{enumitem}

\def\BibTeX{{\rm B\kern-.05em{\sc i\kern-.025em b}\kern-.08em
    T\kern-.1667em\lower.7ex\hbox{E}\kern-.125emX}}
\begin{document}

\title{From Spoofing to Trust: Emergency Alerts \\ Spoofing Testbed and Cross-Cell Verification}

\author{
    \IEEEauthorblockN{Abdallah Abou Hasna\IEEEauthorrefmark{1}\IEEEauthorrefmark{2}, Nada Chendeb\IEEEauthorrefmark{2}, Ammar El Falou\IEEEauthorrefmark{1}}
\IEEEauthorblockA{\IEEEauthorrefmark{1}CEMSE Division, King Abdullah University of Science and Technology (KAUST), Saudi Arabia}
\IEEEauthorblockA{\IEEEauthorrefmark{2}Electrical and Electronics Department, Faculty of Engineering, Lebanese University, Lebanon\\}
Email: \{abdallah.abouhasna, ammar.falou\}@kaust.edu.sa, nchendeb@ul.edu.lb}

\maketitle

\begin{abstract}
Public warning systems (PWS) in cellular networks enable authorities to broadcast emergency alerts to all mobile phones in a geographic region in the event of threats such as earthquakes or severe weather. If an attacker can imitate these alerts and transmit a forged warning containing fake news or phishing links, the impact could range from public panic to user compromise. In this work, we present the first open-source 5G emergency alert spoofing attack, implemented by modifying the openairinterface (OAI) radio access network (RAN) code and executed using a software-defined radio, complemented by a custom network management system to automate network and warning configuration. We conduct a detailed analysis of how different smartphones behave under various conditions. Our findings show that while devices readily display spoofed alerts, the alerting mechanism enables multiple practical attack scenarios beyond simple warning display. Finally, to address this threat, we propose and implement a lightweight cross-cell verification mechanism in OAI, in which the device compares the received warning with neighboring cell broadcasts to flag single-source alerts as suspicious.
\end{abstract}





\section{Introduction}
Public warning systems (PWS) were standardized by the 3rd generation partnership project (3GPP) to provide a unified framework for national authorities to broadcast emergency alerts to mobile users~\cite{3gpp-22.268}. As each country has its own requirements and alert use cases, PWS is designed to deliver different kinds of warning messages, e.g., the commercial mobile alert system (CMAS) in the United States and the earthquake and tsunami warning system (ETWS) in Japan. In 5G networks, these messages are disseminated through system information block (SIB) messages, which are periodically broadcast by base stations to reach all nearby devices.

To ensure maximum reach, 3GPP specifications require emergency alerts to be deliverable without relying on prior authentication or valid subscription~\cite{3gpp-22.268}. Additionally, SIB messages conveying warning information are broadcast without verification or acknowledgment~\cite{5g-spoofing}. Although this choice simplifies dissemination, it also means that any adversary capable of creating a fake base station can impersonate a legitimate network and broadcast forged alerts~\cite {4g-spoofing, 5g-spoofing}. Such spoofed warnings can cause public panic, spread misinformation, or deliver phishing links, especially in such situations where users are under stress and less likely to verify message authenticity.

The first academic experiment on emergency alert spoofing was conducted in~\cite{4g-spoofing}, which demonstrated the feasibility of the attack in 4G networks. Recently,~\cite{5g-spoofing} reproduced the same attack on 5G networks. However, they relied on a commercial product (Amarisoft), which introduces additional cost and limits implementation flexibility to try different attack scenarios and variations. Additionally, existing discussions of defenses remain mostly high-level, with limited validation through real open-source implementations.

In this work, we revisit emergency alert spoofing from a practical and fully reproducible perspective. We implement the transmission of 5G standalone (SA) warning messages using an open-source platform, specifically the openairinterface (OAI) project~\cite{oai}, along with a custom network management system (NMS) serving as a high-level user interface for configuring network parameters. This fully controllable setup allows us to investigate different attack scenarios that prior works did not focus on, including edge cases related to warning size, segmentation, and parallel message handling. We also perform the first systematic analysis of how modern smartphones parse the content of these warnings, including uniform resource locators (URLs), multi-language messages, and other complex text patterns.

Beyond analyzing the attack surface, we address the mitigation problem with a simple and effective defense. Instead of relying on generic recommendations, we propose a cross-cell verification method in which the user equipment (UE) checks whether the same warning is also broadcast by neighboring cells within a short time window. We implement and evaluate this mechanism in OAI, allowing us to study both the attack and the mitigation within the same experimental testbed.

In brief, this paper makes the following contributions:
\begin{itemize}
    \item We present the first fully open-source implementation of a 5G SA emergency alert spoofing attack, extending the OAI project to broadcast valid warning messages.
    \item We develop the NMS interface that automates parameter changes to the public land mobile network (PLMN) and the generation of warning messages, making emergency alert spoofing attack easily reproducible and extensible.
    \item We conduct a detailed study of how smartphones parse and handle warning messages under different conditions.
    \item We propose and implement a cross-cell verification mechanism in OAI as a lightweight and backward-compatible countermeasure against spoofed warning broadcasts.
\end{itemize}

\section{Background}
\subsection{Protocol and Warning Message Flow}
The PWS relies on the cell broadcast service (CBS) to disseminate emergency alerts to mobile users~\cite{3gpp-23.041}. At a high level, an external information source, referred to as the \textbf{cell broadcast entity} (\textbf{CBE}) and operating outside the 3GPP scope, generates warning information, including the alert text, the affected area, and the time period. This information is delivered to the \textbf{cellular broadcast center function} (\textbf{CBCF}).

The \textbf{CBCF}, based on the target area, forwards the warning request to the \textbf{access and mobility management functions} (\textbf{AMFs}) serving the impacted region. Each \textbf{AMF} then forwards the request to the corresponding gNodeBs (gNBs), which repeatedly broadcast the warning over the air on the impacted cells. The UE, finally, receives the broadcast message and displays the warning to the user. A detailed description of this procedure is provided in~\cite{3gpp-23.041}.

While this architecture involves multiple core network components, our work focuses exclusively on the radio access part between the gNB and the UE. The attack we study targets the broadcast nature of warning messages at the air interface and is therefore independent of the internal signaling between the \textbf{CBE}, \textbf{CBCF}, and \textbf{AMF}.

\subsection{Warning Message Delivery in RAN}

In 5G, SIBs are broadcast messages periodically transmitted by the gNB to all UEs within the cell~\cite{3gpp-38.331}. They convey general information about the network, including cell configuration, scheduling parameters, and warning notifications. Among these, \textbf{SIB1} plays a critical role, as it provides essential information to connect to the network, and the scheduling information that allows UEs to locate other SIBs. \\SIBs other than SIB1 are carried inside system information (SI) messages, each of which may carry one or multiple SIBs.

Emergency alerts are delivered using different types of SIBs. \textbf{SIB6} and \textbf{SIB7} are reserved for ETWS notifications in Japan for earthquake and tsunami warnings. \textbf{SIB8} was first used to deliver CMAS messages in the US, but it has been adopted by multiple regions for different alert schemes, including EU-Alert in Europe~\cite{3gpp-02.900} and the Korean public alert system (KPAS). In this work, we focus on SIB8 due to its general applicability. However, the underlying mechanisms discussed also apply to SIB6 and SIB7.

\begin{figure}
  \centering
  \includegraphics[width=\linewidth]{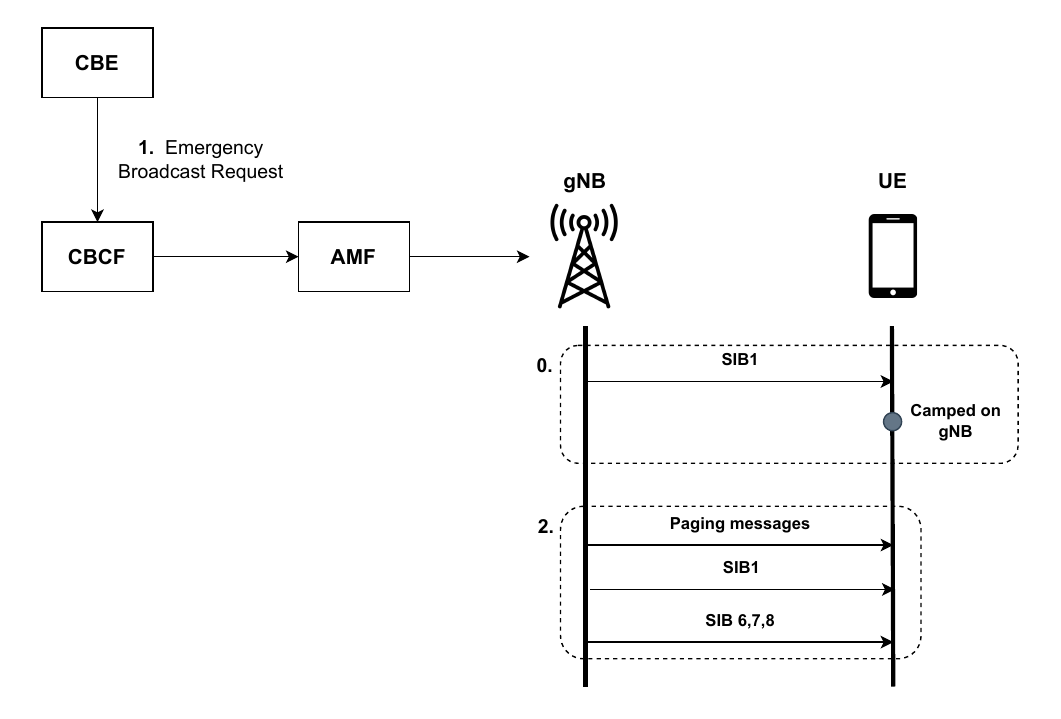}
  \caption{Warning Message Overall Delivery Procedure.}
  \label{fig:warning-delivery}
\end{figure}

Upon the initiation of a warning message, UEs are notified via \textbf{paging} that a new SI message is available~\cite{3gpp-38.331}. The UE subsequently reacquires SIB1 (i.e., re-decodes the new SIB1 to obtain updated scheduling information), then uses this information to acquire SIB8 and parse the warning content. Fig.~\ref{fig:warning-delivery} summarizes the overall delivery procedure.

\subsection{SIB8 Structure}
SIB8 encapsulates warning messages using a set of parameters defined in~\cite{3gpp-23.041,3gpp-38.331}. The most important fields include:
\begin{itemize}
    \item \textbf{Message Identifier}: identifies the warning category. The interpretation of this field depends on the alerting scheme deployed by the country, e.g., alert levels in EU-Alert or presidential alerts in CMAS.
    \item \textbf{Serial Number}: allows UEs to distinguish between old alerts, new alerts, and updates of previously received warnings with the same identifier.
    \item \textbf{Data Coding Scheme}: specifies the encoding used for the warning text. In practice, warnings are commonly encoded using one of the following character encodings:
        \begin{itemize}[nolistsep,noitemsep]
            \item \textbf{GSM $7$-bit}: supports a limited set of characters intended for basic Latin text.
            \item \textbf{UCS2}: uses $16$ bits per character and is required when the warning message contains non-Latin characters or additional symbols, such as Arabic or Cyrillic text.
        \end{itemize}
    \item \textbf{Warning Message Segment Type}: indicates whether the transmitted SIB8 represents the last segment or not.
    \item \textbf{Warning Message Segment Number}: identifies the position of a segment within a multi-segment warning.
    \item \textbf{Warning Message Segment}: carries the warning text, which is organized in a page-based structure\footnote{According to 3GPP specifications, the warning message content follows the cell broadcast message structure~\cite{3gpp-23.041}, where each page contains 82 bytes of user data and a warning message may consist of up to 15 such pages.}.
\end{itemize}
The maximum SI message size is $2,976$ bits ($372$ bytes)~\cite{3gpp-38.331}. As a result, when the warning message segment cannot fit within a single SI message, it must be split across multiple SI messages. Understanding these parameters is essential for correctly crafting SIB8 messages in our implementation, as discussed in Section~\ref{sec:testbed}.

Note that the terms ``alert" and ``warning" are used interchangeably in the literature.

\section{Threat Model and Attack Overview}
Emergency alert spoofing is possible because of a broader cellular network vulnerability: the feasibility of deploying a \textbf{fake base station}.

\subsection{UE Cell Selection and Camping Behavior}

When a UE is powered on or when it exits airplane mode, it initiates the cell selection procedure in order to camp on a cell (i.e., choose a cell and start monitoring its broadcast channels) and enter the \texttt{RRC\_IDLE} state~\cite{3gpp-38.304}. During this process, the UE scans all supported frequency bands, synchronizes with detected cells, and decodes their broadcast channels. By reading SIB1, the UE obtains essential network parameters, including the PLMN.

The UE then selects a suitable cell to camp on, based on radio conditions and PLMN matching, prioritizing cells belonging to the home PLMN (HPLMN)~\cite{3gpp-23.122}.

After camping, the UE continuously monitors neighboring cells to perform cell reselection and may camp on another cell if it detects better radio conditions. Interestingly, this entire procedure is performed using broadcast information and signal strength measurements and does not involve any authentication between the UE and the network. As a result, an attacker operating a fake base station that transmits suitable SIB1 parameters, most importantly a matching PLMN, at sufficiently high power can cause nearby UEs to camp on his cell. This weakness, and its broader security implications, have been discussed in detail in prior works~\cite{park-fakebts}.

\begin{figure}
  \centering
  \includegraphics[width=\linewidth]{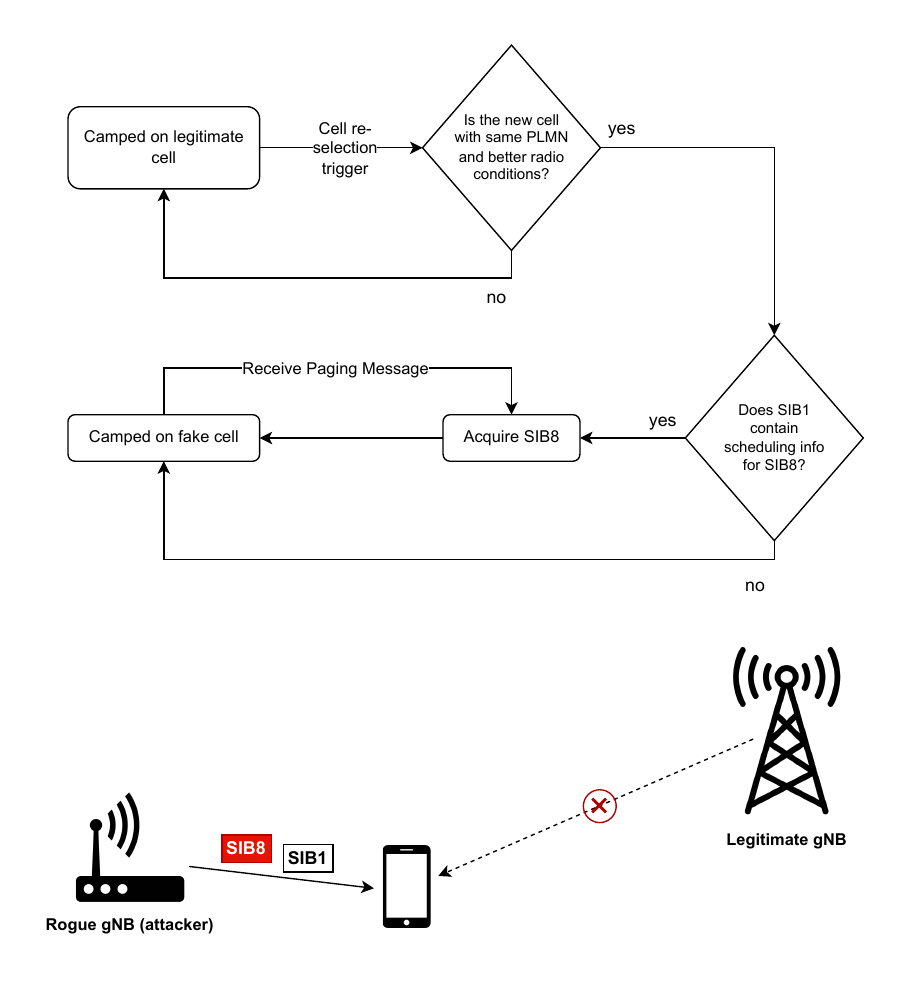}
  \caption{Emergency Alert Spoofing via Cell Reselection.}
  \label{fig:attack-flow}
\end{figure}

\begin{figure}
  \centering
  \includegraphics[width=\linewidth]{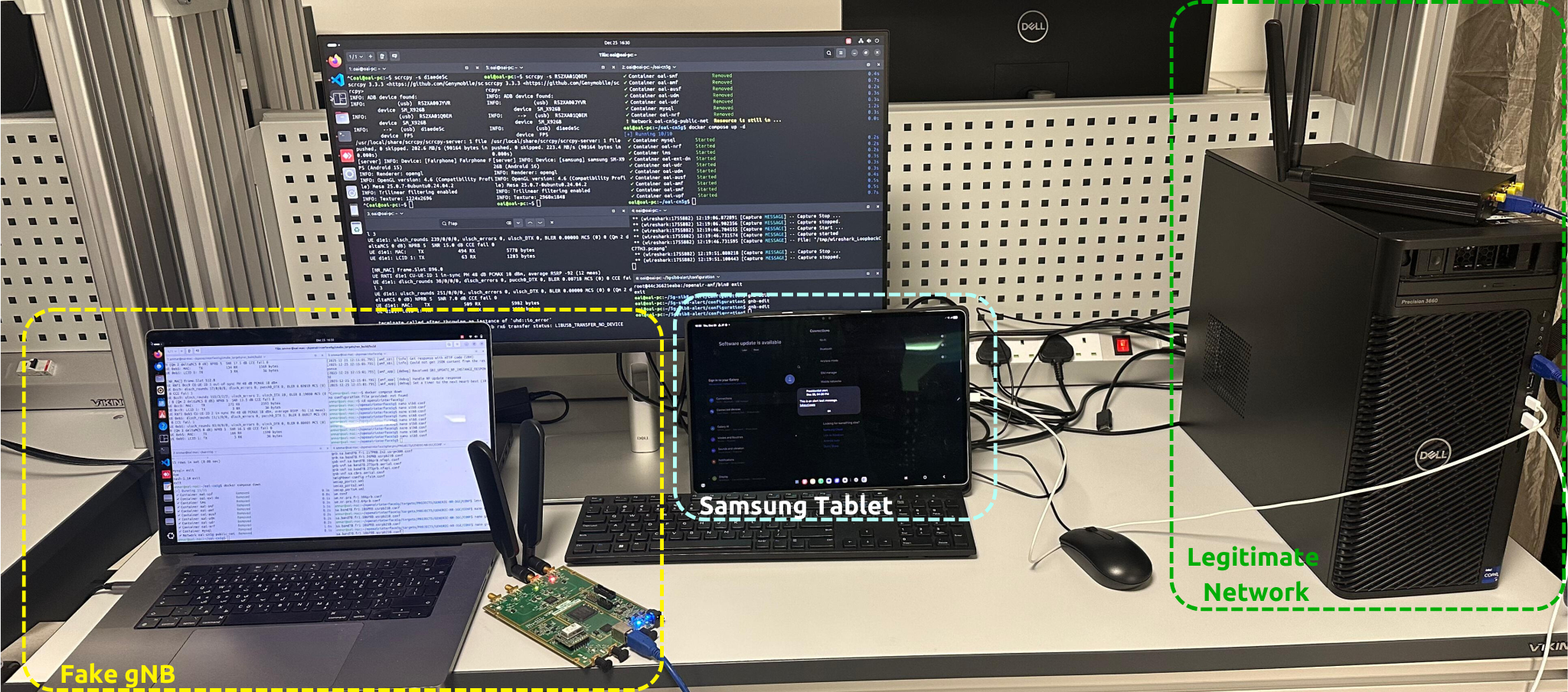}
  \caption{Experimental testbed.}
  \label{fig:testbed}
\end{figure}

\begin{table}
\centering
\normalsize
\caption{Tested devices and operating systems.}
\label{tab:devices}
\begin{tabular}{lll}
\hline
\textbf{Device} & \textbf{OS Version} \\
\hline
Fairphone 5 & Android /e/OS v3.2 \\
Samsung Galaxy Tab S10 Ultra 5G & Android 16 \\
Nothing Phone (3a) & Android 16 \\
Samsung Galaxy A36 5G & Android 16 \\
iPhone 14 Pro & iOS 26.1 \\
\hline
\end{tabular}
\end{table}

\subsection{Warning Delivery in Idle and Inactive States}

3GPP specifications explicitly state that warnings should be delivered to UEs even when they are not connected to the network~\cite{3gpp-22.268,3gpp-38.304}. This means that UEs in the \texttt{RRC\_IDLE} or \texttt{RRC\_INACTIVE} states must be able to receive warning notifications. Therefore, an attacker does not need to establish a full network connection with the UE in order to deliver a fake warning. Once the UE camps on the rogue cell, it will monitor the broadcast channel and acquire SIB8 scheduling information via SIB1. While paging can be used to notify about the availability of new warnings, it is not strictly required if the attacker transmits the scheduling of SIB8 in SIB1 from the beginning (see Fig.~\ref{fig:attack-flow}). By transmitting forged SIB8 messages, the attacker can cause an emergency alert to be displayed without implementing authentication procedures.

Usually, the UE attempts to transition to the \texttt{RRC\_CONNECTED} state and register with the network in order to obtain data services. Since the attacker lacks the cryptographic credentials required for successful authentication, the attempt will fail. However, this process involves multiple retries from the UE; therefore, the warning can be displayed before the UE abandons the rogue cell and reselects another one.

\subsection{Scope and Assumptions}

The attack described in this paper targets UEs in the \texttt{RRC\_IDLE} or \texttt{RRC\_INACTIVE} states. If a UE is already in the \texttt{RRC\_CONNECTED} state, the attacker must first disrupt the connection so that the UE restarts the cell selection procedure. This is because cell reselection occurs without authentication, whereas handover is managed by the legitimate network and would fail for a rogue base station~\cite{4g-spoofing}. This transition can be achieved using simple jamming, which causes the UE to lose connectivity and search for a new cell.

We therefore assume an attacker capable of deploying a rogue gNB to attract idle or inactive UEs in a localized area, a capability already demonstrated in prior fake base-station attacks~\cite{4g-spoofing,5g-spoofing,park-fakebts}.

\begin{figure*}
  \centering
  \includegraphics[width=\linewidth]{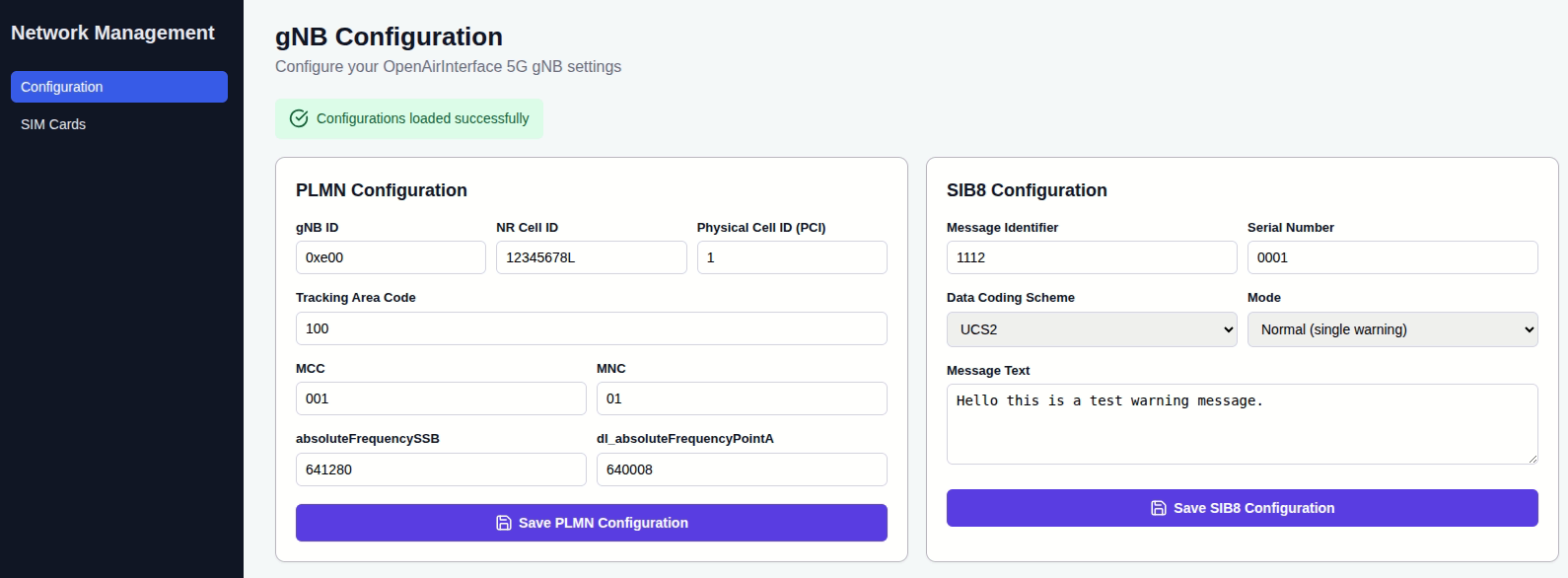}
  \caption{NMS interface for OAI gNB and SIB8 configuration.}
  \label{fig:nms}
\end{figure*}
\section{Testbed and Implementation}
\label{sec:testbed}

\subsection{Experimental Testbed}
Our experimental setup uses a software-defined radio (SDR), specifically the USRP B210~\cite{usrp-b210}, to transmit radio signals generated by the OAI implementation. OAI is an open-source project that is widely used in academia and adopted in a number of commercial deployments. It implements both the radio access network (RAN) and core network functions. Using OAI significantly reduces experimentation cost and, more importantly, enables full control over the attack logic. This was not possible for the prior 5G studies due to the black-box and commercial nature of their experimental platforms.

The testbed is composed of two separate systems. A Dell Precision 3660 desktop running Ubuntu 24.04 LTS is used to operate a legitimate 5G SA network, including both the RAN and the core networks. In parallel, the attacker setup consists of an Apple MacBook with an M1 Pro processor, running Ubuntu 24.04 LTS with a modified OAI gNB without a core network. This attacker configuration highlights the low cost and minimal requirements of the attack, which rely only on a commodity laptop and an SDR.

For end-user evaluation, we tested the attack on multiple commercial smartphones running both Android and iOS. These devices were equipped with programmable SIM cards from Sysmocom~\cite{sysmocom}. Table~\ref{tab:devices} summarizes the tested devices and OS versions. All experiments were conducted inside a controlled environment to prevent interference with real cellular networks. Fig.~\ref{fig:testbed} illustrates the physical testbed setup.

\subsection{OAI Implementation}
The open-source nature of OAI allows fine-grained modifications of gNB behavior, which is essential for implementing and studying our emergency alert spoofing attack.

At the time of our work, OAI did not support broadcasting emergency alerts. We therefore modified the gNB code to enable the generation and transmission of SIB8 messages in compliance with 3GPP specifications. We kept the changes modular to ease future support for disaggregated RAN deployments (centralized and distributed units), for example for drone-based rogue base stations~\cite{el2025base}. 

Warning parameters are read from a configuration file, and SIB8 is constructed within the RRC layer. In order to support realistic warning messages, our implementation allows warning text to be encoded using either GSM 7-bit or UCS2, as specified in~\cite{3gpp-23.038}. We also added runtime support to update the warning parameters while the gNB is running, enabling rapid testing of different alert variants.

In addition, we implemented support for multiple segment transmission, where long warnings are split into multiple SIB8 that share the same message identifier and serial number. 3GPP requires that each SI message carry at most one segment of a given SIB type~\cite{3gpp-38.331}. This requires the transmission of multiple SI messages, along with corresponding updates to the scheduling information in SIB1.

\subsection{Network Management System}

To make our experiments easily reproducible and user-friendly, we developed a lightweight NMS that exposes the main gNB and warning message parameters through a simple web interface, without requiring manual edits to the OAI configuration files. The NMS is containerized, enabling consistent deployment across machines. 

Fig.~\ref{fig:nms} shows the NMS interface. On the warning side, the user can configure SIB8 parameters, including predefined warning transmission modes to automate the scenarios evaluated in Section~\ref{sec:evaluation}. On the gNB side, the user can configure gNB parameters such as the PLMN identity, tracking area code, physical cell identifier (PCI), and frequency-related fields to target a specific operator. For legitimate network use cases, the NMS provides basic SIM management when a core network is used (add, edit, and remove subscriber entries). Overall, the NMS enables researchers to easily reproduce the attack across different test cases and parameter settings in a systematic way.\\The full implementation, including the NMS source code and the OAI patch to enable SIB8 transmission, is publicly available~\cite{sib8-github}.

\section{Evaluation}
\label{sec:evaluation}

\subsection{Warning Reception}

Across all tested devices, forged warnings were received immediately after the UE camped on the rogue gNB. In all cases, the alert was displayed even before the gNB logs indicated the initiation of the random access channel (RACH) procedure. This confirms that neither paging nor successful network authentication is required for warning delivery.

As a result, the core network is not required for this attack. By running only the gNB, fake SIB8 messages were still correctly processed by the UEs.

\subsection{PLMN Configuration}

We evaluated the impact of PLMN identity by changing the PLMN broadcast by the rogue gNB.

When the PLMN of the fake base station matched the UE's HPLMN, all tested devices immediately camped on the rogue gNB and displayed the alert. This behavior is consistent with the normal cell selection and reselection procedures.

When the PLMN did not match the HPLMN, the alert was still displayed on all tested phones after a delay, provided that no cells belonging to the HPLMN were available. According to 3GPP specifications~\cite{3gpp-38.304}, if a UE fails to camp on a suitable cell, it may temporarily camp on any acceptable cell of any PLMN with a high signal quality. In our setup, the rogue gNB satisfied this condition, allowing the phone to eventually camp on it and receive the warning.

\subsection{Parsing of Warning Message Content}

To understand how warning message content is parsed, we tested different types of text in the warning message.

\subsubsection{Web URLs}

Web URLs were recognized as clickable links when they began with an explicit protocol, such as \texttt{http://}, \texttt{https://}, \texttt{rtsp://}, or \texttt{ftp://}. When a protocol was present, all other components of the URL were optional, including user information, the domain name or IP address, the port number, and the path or query string.

If there was no protocol, the URL was still recognized only if it ended with a valid top-level domain (TLD), such as \texttt{.com} or \texttt{.org}. In this case, a minimum of one character before the TLD was mandatory for link detection.

From an attack perspective, URL shortening services increase the effectiveness of phishing attempts. Shortened URLs (e.g., TinyURL) were consistently detected as clickable by all tested devices, while hiding the final destination and any suspicious characters. This removes any visual cue for assessing the legitimacy of links in emergency alerts. Additionally, the Samsung devices and Nothing Phone detected URLs containing Cyrillic letters as valid and clickable, allowing attackers to craft deceptive links that visually mimic legitimate domains.

\subsubsection{Other Clickable Content}

Beyond web URLs, additional content was recognized. Email addresses were detected and the default email application was opened. Phone numbers were also recognized and opened the device dialer.

Map addresses were recognized only when they followed U.S. style, and they were linked to the default map application.

\subsubsection{Language and Encoding}

\begin{figure}
  \centering
  \begin{subfigure}{0.48\linewidth}
    \centering
    \includegraphics[width=\linewidth]{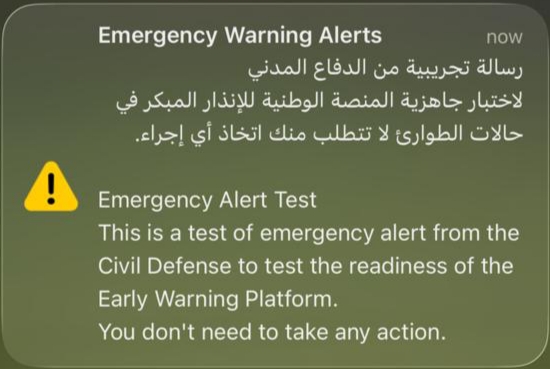}
    \caption{Real public alert test.}
    \label{fig:saudi-real}
  \end{subfigure}
  \hfill
  \begin{subfigure}{0.48\linewidth}
    \centering
    \includegraphics[width=\linewidth]{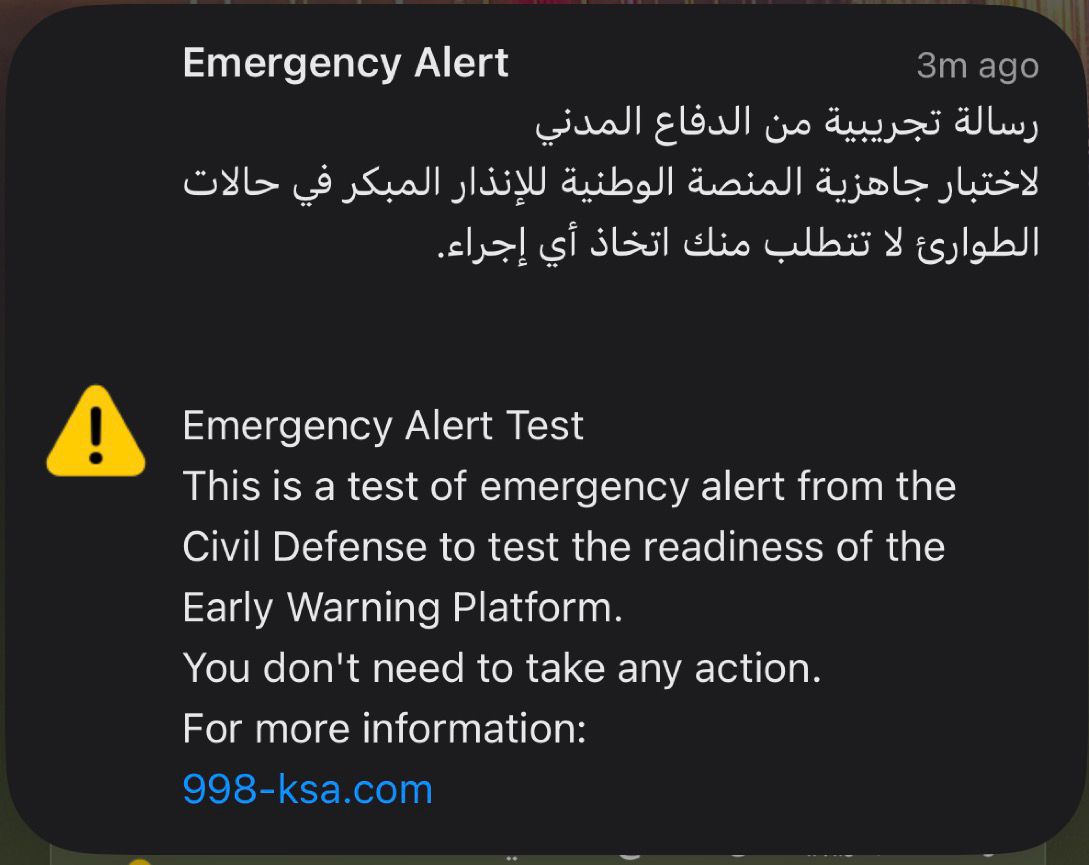}
    \caption{Spoofed alert.}
    \label{fig:saudi-spoofed}
  \end{subfigure}
  \caption{Real vs. spoofed emergency alerts. The attacker can closely mimic the official alert and embed a clickable link.}
  \label{fig:real-vs-fake}
\end{figure}

We confirmed that a single alert can contain mixed-language content, including Arabic and English. UCS-2 encoding supports the use of non-Latin characters and symbols within the basic multilingual plane (BMP), which is important for crafting realistic messages in multilingual regions. To evaluate how closely an attacker can mimic a real warning, we reproduced a real public test alert received in Saudi Arabia and generated a spoofed version with the same bilingual content. Fig.~\ref{fig:real-vs-fake} compares the two alerts, where the attacker additionally embeds a clickable link. 

We also tested whether the alert parser interpreted markup and active content. We found that neither HTML nor JavaScript was rendered or executed. Similarly, lightweight markup formats such as headings (e.g., \texttt{\#}) or bullet-style syntax (e.g., \texttt{-}) were displayed as plain text.

\subsection{Alert Interaction on Locked Devices}

We also evaluated alert interaction when the device was locked. When a forged alert contained a clickable link, and the user selected it, the device prompted for an unlock. However, once unlocked, the link was opened immediately without requiring additional confirmation. This behavior increases the effectiveness of phishing attacks, as users may unintentionally click the link while attempting to open the phone.

\subsection{Segmentation and Serial Number}

We evaluated the handling of segmented SIB8 messages that exceed the size of a single SI message. While most tested devices correctly displayed alerts, the Samsung tablet consistently failed to display alerts with multiple segments, suggesting a lack of segmentation support. We further attempted to broadcast very large warning messages, but messages larger than 15 pages were not displayed on all tested devices.

Next, we tested the reception of multiple single-segment warning messages by transmitting several SIB8 with different serial numbers, up to a maximum of $32$ SI messages~\cite {3gpp-38.331}. All tested devices successfully displayed all the warnings, except the Samsung tablet, which displayed only the last message. Using our runtime update mechanism, we kept updating the broadcast warning by incrementing the serial number during the short period where the UE remains camped on the rogue cell. The repeated alert sounds and vibrations resulted in a highly disruptive user experience.

Finally, we evaluated the handling of segmented warning messages transmitted in parallel. According to 3GPP specifications, when segments belonging to different warning messages (i.e., different message identifiers or serial numbers) are received, a UE may start reassembling them simultaneously, with the number of concurrent reassemblies left to UE implementation~\cite{3gpp-38.331}. To do so, we interleaved segments from several warning messages. In our experiments, devices that support segmentation displayed only a single completed warning, indicating that their implementations limit the number of parallel warning reassemblies.

Overall, these experiments show that emergency alert spoofing involves more than simply displaying a forged warning with fake news or a phishing link. Several behaviors related to warning size, segmentation, and message handling can be influenced in practice, and some depend on device-specific implementations. As a result, similar attacks may have different effects across devices, motivating the need for careful evaluation and appropriate countermeasures.

\begin{figure}
  \centering
  \includegraphics[width=0.96\linewidth]{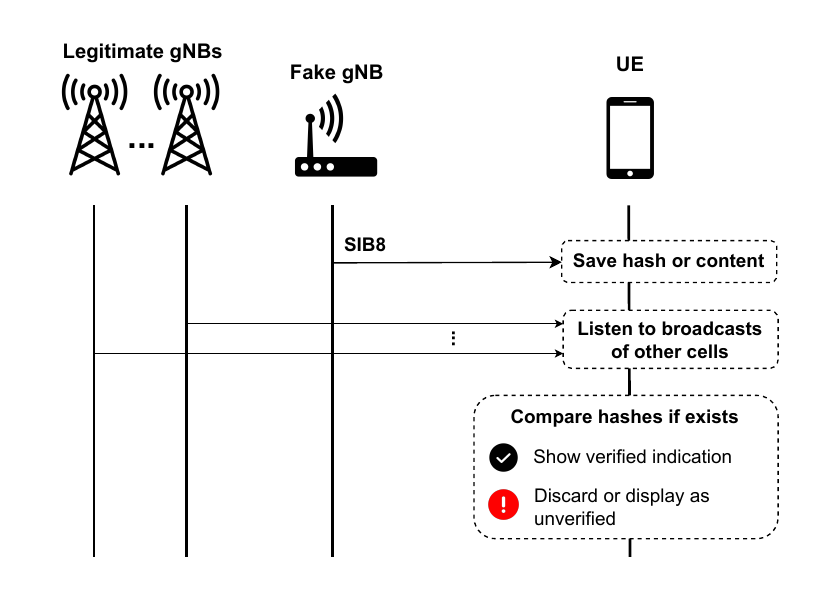}
  \caption{Cross-cell verification workflow.}
  \label{fig:ccv}
\end{figure}

\section{Countermeasures}
Several countermeasures have been proposed in prior work to address emergency alert spoofing, including adding authentication to warning messages or verifying them through external entities~\cite{5g-spoofing, 4g-spoofing}. However, these solutions are hard to deploy in practice. Some require changes to the standard or modifications at the network level. Others depend on external verification, which adds delay and extra traffic. In addition, most of them have not been validated through open and reproducible implementations.\\In this section, we propose a lightweight and practical mitigation technique that is backward compatible and does not require any network-side modification. 

\subsection{Cross-Cell Verification Approach}
The idea is based on a simple observation: legitimate warning messages are typically broadcast periodically by multiple base stations in the affected area, often across different operators, while a spoofed warning is often transmitted by a single rogue base station. By comparing broadcast channels across neighboring cells, the device can assess the legitimacy of the received warning and detect single-source warnings. \\After receiving an emergency alert while in \texttt{RRC\_IDLE} mode, the UE performs the verification steps illustrated in Fig.~\ref{fig:ccv}:

\begin{enumerate}
    \item The UE stores the alert content and associated parameters,
    \item The UE listens to broadcast channels from one or more neighboring cells. When available, cells from different operators are especially useful because their coverage areas usually overlap. This step does not require any network connection or authentication.
    \item The UE checks whether the same warning is being broadcast by other nearby cells within a short time window. The comparison can rely on a warning hash or direct content matching.
    \item If the alert is observed from multiple gNBs, the UE classifies it as verified. If the alert is only observed from a single cell while neighboring cells remain silent, the UE classifies it as unverified.
\end{enumerate} 

Our method is passive and relies only on broadcast system information, without requiring any connection establishment or a valid subscription. It introduces only a minimal delay and can be integrated with limited UE-side changes, since required capabilities, such as neighbor measurements and cell reselection, are already supported on UEs.

\subsection{Implementation}
We implemented the proposed cross-cell verification mechanism directly in the OAI UE code.
While an implementation on commercial phones would be more realistic, it is difficult in practice due to limited access to low-level cellular functions. Achieving this would typically require kernel-level modifications, which is a complex effort, or the use of a privileged application that still does not provide full control over alert handling.

Our goal is to build a proof of concept (PoC) with full control over UE behavior, therefore we relied on the OAI UE. This can serve as a basis for future implementation at the modem or kernel level by device vendors.

The implementation starts at the RRC layer. When the UE decodes an SIB8 message, it does not immediately mark the alert as trusted. Instead, it stores the serving cell PCI together with a hash of the SIB8 content using SHA-256.
\\The RRC layer then triggers a dedicated verification mode in the UE MAC and PHY layers. In this mode, the UE remains in idle-like behavior and does not perform any connection establishment. Instead, it scans a list of carriers and synchronizes to neighboring cells one by one. Already inspected PCIs are excluded to avoid rescanning the same cell. For each detected cell, the UE attempts to acquire system information and checks whether an SIB8 is scheduled. If SIB8 is present, its content is hashed and compared against the hash of the original warning. 

In the current implementation, the verification process is fully configurable. The user can adjust the maximum number of cells to scan, how many matching SIB8 messages are needed to consider that the warning is verified, and the duration of the scan. If enough matching messages are found, the alert is marked as verified. Otherwise, if this condition is not met after scanning the selected cells or when the scan timer expires, the alert is marked as unverified.\\The implementation of the cross-cell verification mechanism is also included in our publicly available repository~\cite{sib8-github}.

\subsection{Evaluation}
We evaluated the proposed mechanism using our OAI-based testbed, where the OAI UE is connected to a separate SDR (USRP B210), allowing it to independently scan and synchronize with cells during the verification process.

In our experiments, the UE triggers the verification procedure right after receiving an SIB8 message and proceeds with scanning neighboring cells. The behavior was as expected: when the number of matching warnings reaches the configured number, the alert is marked as verified. Otherwise, the UE flags the warning as unverified. In isolated scenarios where no neighboring cells were available, the UE was unable to complete the verification process and consequently marked the warning as unverified.

\section{Conclusion}
In this work, we examined the security of 5G emergency alert broadcasting from a practical perspective. We implemented the first open-source 5G warning spoofing attack by extending the openairinterface project to broadcast forged SIB8 warning messages using an SDR. We complemented this testbed with a lightweight, containerized management system that automates gNB and warning configuration, enabling fast and reproducible experiments. Our experiments show that once a phone camps on a rogue cell, emergency alerts can be displayed even without a core network. Beyond simple alert display, we showed that, in practice, warning handling can expose additional attack scenarios that vary across devices. Finally, we proposed a lightweight cross-cell verification mechanism at the UE side to detect single-source warnings. We implemented and evaluated this approach in OAI, showing that it can effectively distinguish between isolated fake warnings and legitimate multi-cell alerts.

\bibliographystyle{IEEEtran}
\bibliography{sample-base} 

\end{document}